\documentclass[%
 reprint,
 amsmath,amssymb,
 aps,
floatfix,
]{revtex4-2}

\usepackage{graphicx}% Include figure files
\usepackage{dcolumn}% Align table columns on decimal point
\usepackage{bm}% bold math
\usepackage{color}
\usepackage{physics}
\usepackage[normalem]{ulem}

\begin{document}

%\preprint{APS/123-QED}

\title{The Multi-terminal Inverse AC Josephson Effect}

\author{Ethan G. Arnault$^{1\ast}$, Trevyn Larson$^{1}$, Andrew Seredinski$^{1,2}$, Lingfei Zhao$^{1}$,
\\
Sara Idris$^{3}$, Aeron McConnell$^{3}$, Kenji Watanabe$^{4}$, Takashi Taniguchi$^{4}$, Ivan V. Borzenets$^{5,6}$,
\\
Fran\c{c}ois Amet$^{3}$, Gleb Finkelstein$^{1}$
\\
\normalsize{$^{1}$Department of Physics, Duke University, Durham, NC 27708, USA}
\\
\normalsize{$^{2}$School of Sciences and Humanities, Wentworth Institute of Technology, Boston, MA 02115, USA}
\\
\normalsize{$^{3}$Department of Physics and Astronomy, Appalachian State University, Boone, NC 28607, USA,}
\\
\normalsize{$^{4}$Advanced Materials Laboratory, NIMS, Tsukuba 305-0044, Japan
}
\\
\normalsize{$^{5}$Department of Physics, City University of Hong Kong, Kowloon, Hong Kong SAR}
\\
\normalsize{$^{6}$Department of Physics and Astronomy, Texas A\&M University, College Station, Texas 77843, USA}
\\
\normalsize{$^\ast$To whom correspondence should be addressed; E-mail:  ethan.arnault@duke.edu}
}

\date{\today}% It is always \today, today,

\begin{abstract}
When a Josephson junction is exposed to microwave radiation, it undergoes the inverse AC Josephson effect -- the phase of the junction locks to the drive frequency. As a result, the $I-V$ curves of the junction acquire  ``Shapiro steps'' of quantized voltage. If the junction has three or more superconducting contacts, coupling between different pairs of terminals must be taken into account and the state of the junction evolves in a phase space of higher dimensionality. Here, we study the multi-terminal inverse AC Josephson effect in a graphene sample with three superconducting terminals. We observe robust fractional Shapiro steps and correlated switching events, which can only be explained by considering the device as a completely connected Josephson network. We successfully simulate the observed behaviors using a modified two-dimensional RCSJ model. Our results suggest multi-terminal Josephson junctions are a playground to study highly-connected nonlinear networks with novel topologies.
\end{abstract}

\maketitle

\section{Introduction}

The superconducting phase of a resistively and capacitively shunted Josephson junction (JJ) has the same dynamical properties as a pendulum \cite{Stewart1968,McCumber1968}. This endows Josephson junctions with a rich phenomenology at the confluence of nonlinear and quantum dynamics \cite{KautzRev1996}. Networks of JJs in particular offer the opportunity to design a highly-tunable nonlinear oscillator in arbitrary dimension. Indeed, Josephson arrays have been shown to host not only a variety of quantum states, but also  nonlinear dynamical behaviors such as synchronization \cite{StrogatzSynch}, chimera states \cite{Chimera}, splay states \cite{Splay1,Splay2}, and chaos \cite{Stroud89}.

While Josephson arrays have been studied extensively (see e.g. Refs. \cite{Martinoli2000, Tinkham, JJArrayFrac}), the phase dynamics of driven multi-terminal Josephson junctions has yet to be fully explored. In multi-terminal junctions, a Josephson coupling is established between each pair of superconducting contacts across a common normal channel (for instance, graphene). An electron microscope image of such a device is shown on Figure 1a in the three terminal case. The added complexity makes multi-terminal junctions an ideal medium for engineering novel quantum and topological phenomena. For example, the energy spectrum of multi-terminal Josephson junction based on a few-mode semiconductor has been predicted to emulate the band structure of topologically non-trivial materials \cite{Riwar2016,Eriksson2017,Meyer2017,Xie2017,Xie2018,AkhmerovMTJJ,WeylCircuit,Gavensky2018,Venitucci2018,GeomtricTensor}. This exciting prospect led to renewed efforts toward experimental realizations of multi-terminal Josephson junctions~\cite{Pfeffer2014,Strambini2016,Cohen2018,Draelos2019,Pankratova2020,Pribiag2020}, which calls for new insights into their phase dynamics.

\begin{figure}[htp]
    \centering
    \includegraphics[width= .75\columnwidth]{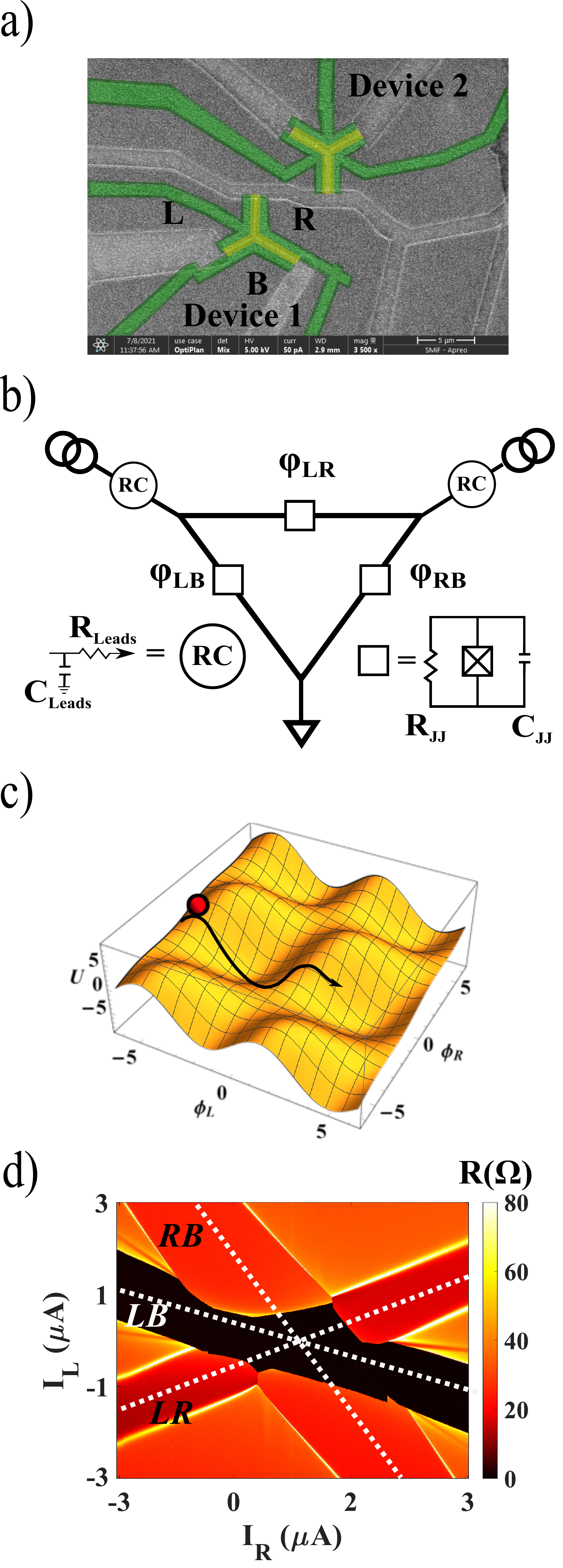}
 \caption { a) An SEM image of the studied device. A 500 nm channel of monolayer graphene is contacted by three MoRe leads of widths between 6.5 and 7.5 $\mu$m. Current $I_L$ is sourced from $L$ and $I_R$ from $R$. $B$ is grounded. Voltages are measured between each contact and ground. b) A circuit model of the device. All three superconducting contacts are connected via Josephson junctions, resistors and capacitors. c) An illustration of the trajectory of the phase particle through the two-dimensional washboard potential. d) The differential resistance (dV$_{LB}$/dI$_L$) bias-bias map at zero microwave power and Vg = 10 V. The three branches corresponds to supercurrent between three pairs of contacts.}
    \label{fig:Fig0}
\end{figure}

An important step in characterizing junction dynamics is exploring the evolution of the superconducting phase under a microwave excitation~\cite{JosephsonRMP1964}. In a conventional two terminal Josephson junction, the junction phase could lock to the phase of the external drive. This results in Shapiro steps: the rectification of the applied microwave current which generates a quantized DC voltage across the junction $V=\frac{\hbar}{2e}\left< \frac {d \phi}{dt} \right> $ at integer or fractional multiples of $V=\frac{\hbar \omega}{2e}$, where $\omega$ is the drive frequency \cite{Shapiro1963}. The phase locking is well understood in the context of the Stewart-McCumber (RCSJ) model, whereby an imaginary phase particle rolls down the rungs of a tilted washboard potential \cite{Tinkham}. Since the washboard potential is determined by the junction's current-phase relationship (CPR), Shapiro steps can serve as a probe into the CPR, which is especially useful for studying Josephson junctions made from unconventional materials \cite{Kwon2004,FuKane2009,Rokhinson2012,Wiedenmann2016,quasiparticleHeating,Platero2017}.

The state of the three-terminal junction is described by two independent phases, e.g. the phases of the left ({\bf L}) and right ({\bf R}) terminals with respect to the grounded bottom ({\bf B}) terminal (Figure 1a,b). The two phase differences $\phi_{LB}$ and $\phi_{RB}$ encode a washboard potential, as illustrated on Figure 1c and detailed in the supplementary. The applied DC currents tilt the washboard potential in a certain direction, while an AC excitation generates a rocking motion. In a general case, the AC currents applied to the two contacts could  differ both in amplitude and in phase. The resulting out-of-phase rocking of the washboard potential along each axis therefore tends to push the phase along elliptical trajectories, which become open, drifting spirals if DC biases are superimposed to the periodic drive. 

In this paper, we unravel the phase dynamics of a three-terminal Josephson junction under microwave irradiation. In this case, mode-locking can occur both between the drive and the two phases $\phi_{LB}$ and $\phi_{RB}$, as well as between the two phases themselves due to their coupling. Indeed, we observe signatures of collective behaviors, such as fractional phase locking and oscillator synchronization. We qualitatively model the system with a multi-terminal generalization of the conventional RCSJ model \cite{Tinkham}.

\begin{figure}[htp]
    \centering
    \includegraphics[width= .95\columnwidth]{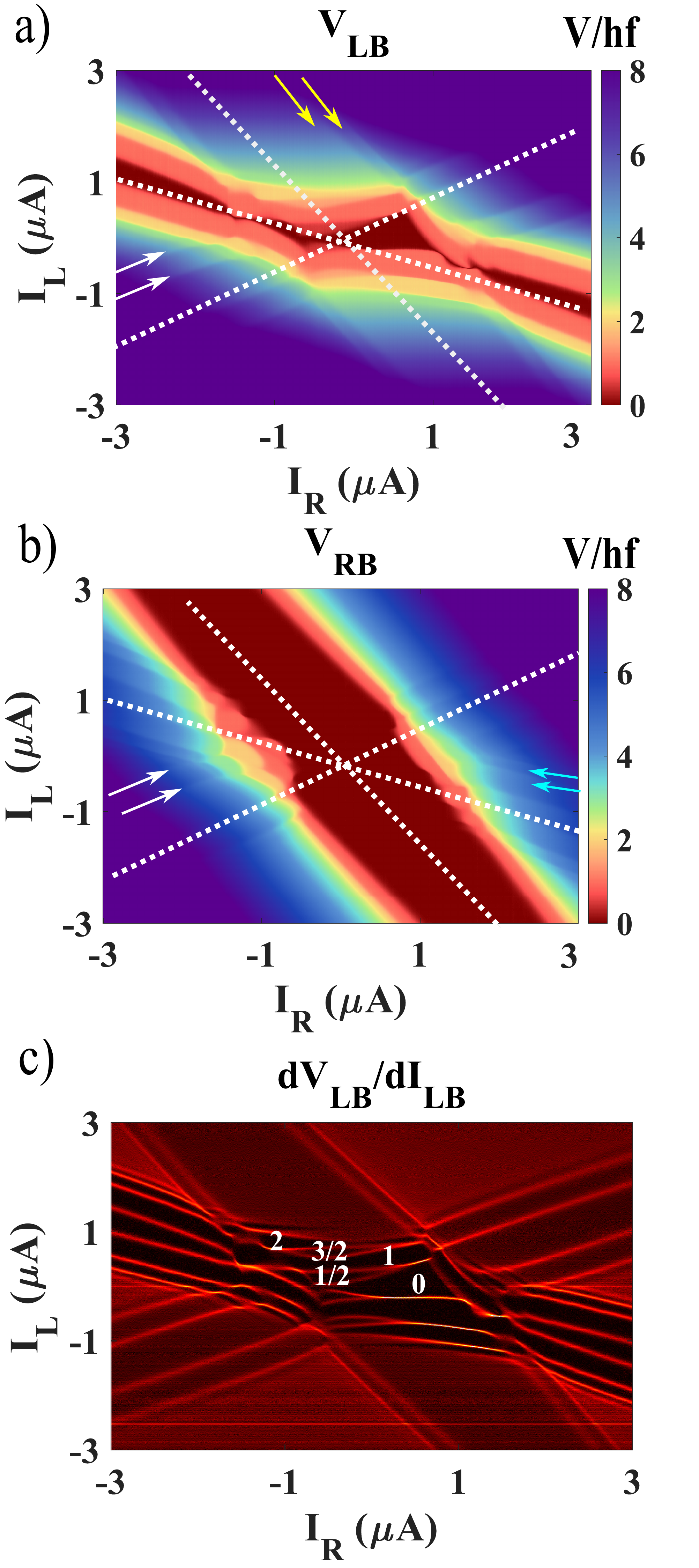}
    \caption {DC voltage maps for junctions $LB$ (a) and $RB$ (b) at 5.2 GHz, 15 dBm, and $V_{bg} = 10$ V. White dashed lines indicate the biasing conditions which result in zero voltage across one of the three junction. At high bias, imprints of other junctions emerge as ripples on each map parallel to the white dashed lines (arrows mark the location of the ripples, cyan from $LB$, yellow from $RB$, white from $RL$). c) Differential map of panel a to highlight the plateau boundaries. Half integer Shapiro steps emerge in the $LB$ (and $LR$) junction at the intersection of the ripples induced by the $RB$ junction. These half integer steps disappear for high applied biases, where only the single junction behavior is present.
    }
    \label{fig:Fig1}
\end{figure}

\section*{Results}

Our multi-terminal junction consists of monolayer graphene encapsulated in hexagonal boron nitride, which protects the device from fabrication contaminants and enables ballistic transport over several microns~\cite{Dean2010, Mayorov2011}, including ballistic supercurrents ~\cite{Borzenets2016, Calado2015,BenShalom2016}. The junction's three contacts are made of sputtered molybdenum-rhenium (MoRe), a superconductor known to form high transparency Ohmic contacts to graphene~\cite{Calado2015}. The primary device studied here is Y-shaped, with the three junctions meeting at a 120 degree angle (Figure 1a). The contacts are separated by 0.5 $\mu$m long graphene channel, and the junctions widths (lateral extent of the contacts) range between 6.5 and 7.5 $\mu$m. Fabrication details are provided in the Supplementary Information.

\begin{figure*}[htp]
    \centering
    \includegraphics[width= 2\columnwidth]{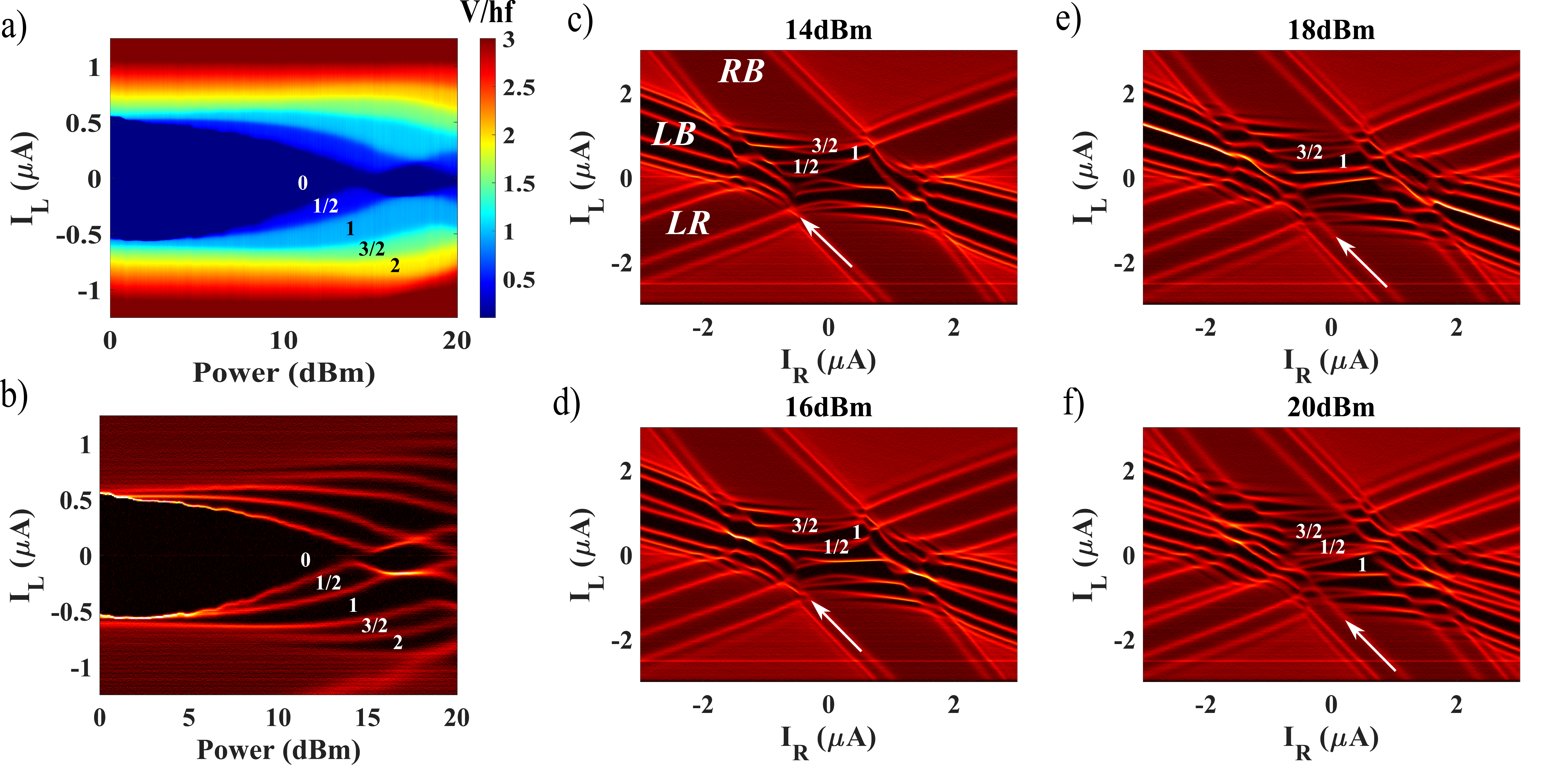}
    \caption {a) $V_{LB}$ at fixed I$_R$ = -300 nA while varying I$_L$ and applied microwave power. The Shapiro steps demonstrate the prototypical Bessel function behavior, including the half-integer steps. Notably, the step width of of the fractions is comparable to the width of the integer steps. b) Numerical derivative of 3a as a guide to the eye. c-f) Differential bias-bias maps tracking the evolution of the plateaus with changing microwave power. At the boundary generated by the $RB$ $\pm$ 1 step, the fractional plateaus are forced to switch (denoted by white arrows). 
    }
    \label{fig:Fig2}
\end{figure*}

The sample is cooled to a base temperature of 50 mK in a Leiden Cryogenics dilution refrigerator. However, additional heating from the microwave signal can warm the sample holder to several hundred mK. In fact, it has been reported that in similar microwave-driven  
SNS junctions the actual electron temperature may be even higher \cite{quasiparticleHeating}. Unless otherwise stated, we apply 10 V to the back gate, which corresponds to an electron density of 2.5$\times  10^{12}$/cm$^2$. A small magnetic field offset of 0.5 mT is used to suppress the critical current of the junctions. This allows us to tune between different nonlinear regimes and suppress hysteresis in our Shapiro steps~ \cite{Trevyn,KautzRev1996}. In the supplementary, we present two representative results measured at zero field and at a different gate voltage. 

In the typical measurement, DC bias currents $I_L$ and $I_R$ are applied to the {\bf L} and {\bf R} contacts with respect to the grounded {\bf B} contact. We present either the DC voltages $V_{LB}$ and $V_{RB}$, or the low-frequency differential resistances $dV_{LB}/dI_{L}$ and $dV_{RB}/dI_{R}$ (measured using a lock-in amplifier). In Figure 1d, we plot $dV_{LB}/dI_{L}$ vs. $I_L$ and $I_R$, measured without applied microwave radiation. The resulting map is consistent with previous studies by us and other groups~\cite{Draelos2019,Pankratova2020,Pribiag2020}. Three superconducting branches emerge corresponding to supercurrents flowing between each pair of contacts. Additional ``quartet'' features emerge between the branches \cite{Pfeffer2014,Cohen2018}, which will not be the focus of this study.

We now send a 5.2 GHz signal to an antenna located $\sim 1$ mm from the device. The microwave excitation applied at room temperature by the signal generator passes through several attenuators, reducing the power reaching the sample by at least 5 orders of magnitude. Since the delivered power varies as a function of frequency, we quote only the power applied by the signal generator. In Figure 2, we apply 15 dBm of microwave power and measure the voltages of each junction. As a guide to the eye, we add dashed white lines, which correspond to contours of constant voltage across all three Josephson junctions $RB$, $LB$ and $LR$, as extracted from Figure 1d (measured in the absence of the microwave excitation). 

When a sufficiently large current bias is applied to a given contact, the corresponding phase is not locked to the drive and its voltage with respect to ground is not quantized. However, another junction in the network may simultaneously exhibit Shapiro steps, which couple to the measured junction through the device's circuit network and cause voltage ripples.
For instance, note the ripples in $V_{LB}$ marked by yellow arrows in Figure 2a, which are parallel to the lines of constant $V_{RB}$ in Figure 2b. This feature corresponds to the imprint of junction $RB$ onto junction $LB$. In a similar fashion, the imprint of $LB$ on $RB$ is observed in Figure 2b (shown by cyan arrows); imprints from $LR$  (not shown) are observed in both maps (shown by white arrows). 

Figure 2c highlights the step boundaries by showing the numerical derivative of the $LB$ voltage map. Here, at the intersection of all three voltage branches, the junctions' mutual influence become even more pronounced, resulting in the appearance of prominent half integer steps. Depending on the parameters, the half integer steps are observed between any pair of contacts (see supplementary).

To elucidate the origin of the half integer steps, we consider several observations: 1) Upon increasing the bias out of the central region ($I_{L,R}<1\mu$A), only integer steps remain - the half integer steps have disappeared. 2) The half steps are washed away at about the same temperature as the integer steps (see supplementary information). 3) In the previous measurements of individual graphene junctions of similar dimension, which used the same experimental setup, we have not observed such prominent fractional steps as in Figure 2 \cite{Trevyn}. As we argue further in the text, these three observations indicate that the origin of the fractional steps is neither due to non-sinusoidality in the CPR \cite{Lubbig1974,FractGraphene,PbSnTe, InAs,Bae2008,Dinsmore2008}, nor the circuit components external to the junction such as a capacitor or inductive load \cite{Sullivan1970,Russer1972}. Instead, the most likely origin of the observed behavior is the multi-terminal nature of the present sample. Indeed, it has been predicted that Josephson networks could demonstrate fractional steps due to the full breadth of their circuitry, even when their constituent junctions have sinusoidal CPR~\cite{TriangularArray}.

To understand the origin of the half-integer plateaus, we have to consider spiraling trajectories in a two-dimensional washboard potential of Figure 1c. These trajectories can cause the phase to land on a saddle point of the washboard potential after a cycle of the drive, rather than a global minimum. It then could take more than one cycle to reach the global minimum, which yields fractional voltage plateaus. This intuitive picture is substantiated by the numerical simulations in Figure 4c and 4d, which we describe later in more detail. Here, the two-dimensional washboard is tilted horizontally by $I_L$, while $I_R=0$. The resulting rocking motion in the $\phi_R$ direction keeps $V_R=0$ on average. However, the trajectory spirals down along the $\phi_L$ direction, progressing by $2\pi$ every two cycles. As a result, a substantial half frequency subharmonic of the applied microwave excitation is generated. The rectification of this $\omega$/2 frequency results in fractional Shapiro steps with $V_L=1/2$ $\frac{\hbar \omega}{2e}$.

We next turn our attention to the evolution of the Shapiro steps with the power of the microwave signal. In Figure 3, $I_L$ is swept, while $I_R$ is tuned to $\sim-300$ nA, which is the offset bias best suited to capture both the $\frac{1}{2}$ and $\frac{3}{2}$ steps (see Figure 2c). To better identify the plateau boundaries, we take a numerical differential of Figure 3a with respect to $I_L$ (Figure 3b).

Two observations are immediately visible in Figure 3a,b: 

First, it is notable that the integer and fractional steps have comparable widths. Shapiro steps at fractional multiples of $\hbar\omega/2e$ can appear in a single junction with either a sinusoidal \cite{Sullivan1970,Russer1972} or a non sinusoidal \cite{Lubbig1974,FractGraphene,PbSnTe, InAs,Bae2008,Dinsmore2008} current phase relation. However, those fractional steps are typically much less robust than their integer counterparts.  On the other hand, frustrated Josephson arrays have demonstrated integer and fractional plateaus of similar widths~\cite{JJArrayGiantFrac}, indicating that network effects in connected Josephson junctions are universally more robust in comparison to individual device dynamics.

Second, both the fractional  and the integer steps in Figure 3a,b all show qualitatively the same behaviour as the integer steps in a prototypical single junction \cite{Tinkham}. There, the plateau widths are described by  $I_cJ_n(\frac{2eV_{AC}}{\hbar \omega})$, where $J_n$ is the Bessel function, which appear very similar to Figure 3. (See e.g. our results in~\cite{Trevyn}.) We are not aware of a simple argument why the half-integer steps in our case would follow the same type of pattern. Moreover, our previous measurements of a single graphene junction of similar dimensions showed strong deviations from the Bessel functions~\cite{Trevyn}. Instead, large overlap between plateaus resulted in pronounced hysteresis which modified the shape of the plateaus. In the supplementary information, we show that when the multi-terminal junction is tuned into the hysteretic regime, the fractional plateaus disappear, being overwhelmed by the integer steps.

\begin{figure*}[htp]
    \centering
    \includegraphics[width= 2\columnwidth]{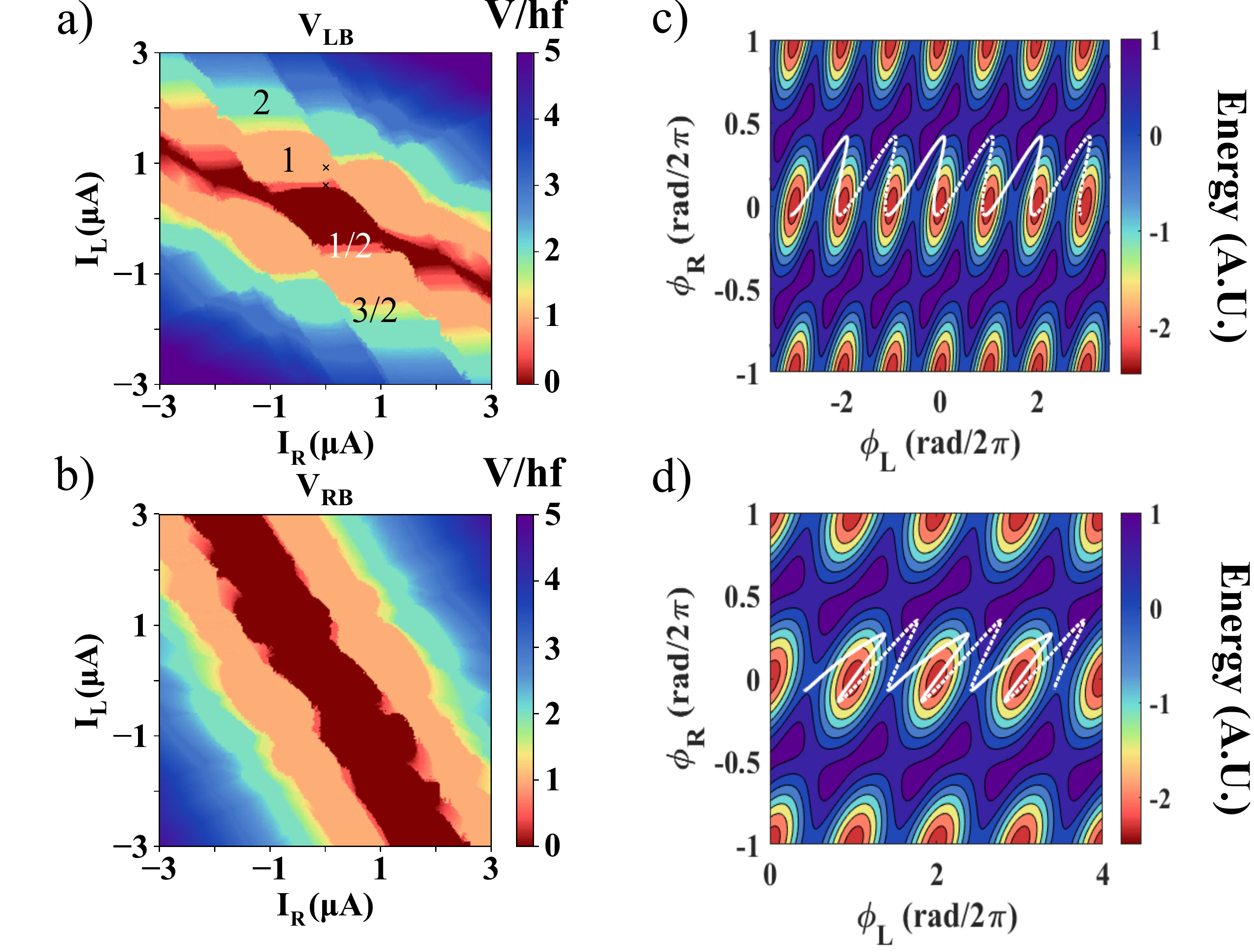}
    \caption{ a,b) Simulated bias-bias maps of the $LB$ (a) and $RB$ (b) junctions. Our simulation produces comparable results to the experiment, capturing both the general shape of the plateaus as well as the fractional steps. Simulation parameters are listed in the supplementary information. X's mark the location in bias where the phase trajectories are taken. c,d) The phase trajectory of the 1 (c) and 1/2 (d) plateaus overlaid atop the energy landscape. Alternating dashed and solid lines correspond to one cycle of the drive. A slow rocking motion develops within the minima. The rocking allows the phase particle to avoid global maxima and run across the diagonal of the potential. This causes a stable cycling finishing near global minima for integers. Meanwhile, trajectories that end near both minima and saddle points generate fractional steps.
    }
    \label{fig:Fig4}
\end{figure*}

The formation of both integer and fractional plateaus is made clear through the study of the bias-bias maps at various microwave powers (Figures 3c-e). At high bias, each junction develops it's own set of integer Shapiro steps. At the same time, in the central region, where the superconducting branches meet, a complicated web of integer and fractional plateaus is formed. 

Focusing on the $\pm 1/2$ steps, we note that the plateaus are interrupted upon reaching the $RB$ junction's $\pm 1$ boundary, which itself is nearly unaffected. (The $RB$ boundary is denoted by the white arrows.) This observation demonstrates that the stability of the fractional steps is weaker than that of the integer steps. We argue that the fractional phase trajectory is critically disrupted by the transition to a running state of the $RB$ junction which traverses multiple periods of $\phi_{RB}$. It is important to note that the variations in dynamical states of the junctions, which fundamentally change the phase trajectory, are more disruptive than the temperature fluctuations for the temperature ranges studied. 

To link our experimental findings and theoretical understanding, we now turn to modelling the dynamics of the multi-terminal junction. We use a fourth order Runge-Kutta method to solve a series of coupled differential equations derived from the multi-terminal generalization of the RCSJ model, schematically shown in Figure 1b. More details, as well as a list of parameters used in the simulation are found in the supplementary information.

Modelling the behavior of Figure 2, we find good qualitative agreement between the simulation and the experiment, capturing both the fractional steps as well as the general shape of the plateaus (Figure 4a,b). The simulations also produce the trajectory of the phase `particle' in the $\phi_{L,R}$ space. Figure 4c,d depict the simulated phase evolution taken at biases corresponding to the V$_{LB}$ = 1 and 1/2 plateaus respectively, while V$_{RB}$ = 0. The trajectories are overlaid atop the energy landscape. In the $\phi_R$ direction, the particle is rocking back and forth to keep
$\left< \frac {d \phi_R}{dt} \right>=0$. Thus, one full cycle of the drive corresponds to $\phi_R$ returning back to the same  value. The rocking allows for the `particle' to climb over regions lower than the typical maxima. The less steep path allows for the particle to traverse several minima in the $\phi_L$ direction before ultimately returning to the same point, shifted $n$ periods away. It is suggestive that for integers, these trajectories seem to end on minima (Figure 4c), while for fractions (Figure 4d), trajectories seem to end on either minima or saddle points. The quasistability of the saddle point may explain why the fractional steps are more dynamically sensitive than the integer steps.

In conclusion, we study the Shapiro steps in a multi-terminal Josephson junction made of graphene, which provides all possible connections between pairs of superconducting terminals. The interplay within the full network of couplings allows for the generation of robust fractional Shapiro steps and correlated transitions between junctions. The fractional plateaus are found to be more sensitive to changes in dynamical state than the integer plateaus, despite having comparable sensitivity to temperature. These findings are simulated using an extended RCSJ model, where it is shown that the fractional plateaus emerge due to nontrivial phase trajectories across the 2D washboard potential.

In the future, adding more contacts will provide unique circuit topologies that are not realizable in two-dimensional Josephson arrays. This may enable experimental observations of chimera \cite{Chimera} and splay states \cite{Splay1,Splay2}, which are predicted in Josephson arrays, but with the added benefit of being able to bias or measure from any junction directly. Additionally, making junctions from novel materials may allow for more complex CPRs, providing a path to previously unexplored dynamical states. Utilizing graphene devices with local gates, one should be able to control the coupling strengths and to change the topology of the system at will. Finally, our observations may lend themselves useful as elements in superconducting microwave circuits. For example, the robust generation of down converted microwave photons, without the need for higher CPR harmonics, may allow for integrated on-chip applications in quantum information processing.

\begin{acknowledgments}
We thank Stephen Teitsworth and Michael Lee for helpful discussions. Transport measurements by E.G.A., F.A. and T.L., lithographic fabrication and characterization of the samples by E.G.A., F.A., A.S., and L.Z., and data analysis by E.G.A., F.A. and G.F., were supported by Division of Materials Sciences and Engineering, Office of Basic Energy Sciences, U.S. Department of Energy, under Award No. DE-SC0002765. S.I. was supported by a GRAM fellowship. F.A., and A. M. were supported by a URC grant at Appalachian State University.  K.W. and T.T. acknowledge support from JSPS KAKENHI Grant Number JP15K21722 and the Elemental Strategy Initiative conducted by the MEXT, Japan. T.T. acknowledges support from JSPS Grant-in-Aid for Scientific Research A (No. 26248061) and JSPS Innovative Areas “Nano Informatics” (No. 25106006). This work was performed in part at the Duke University Shared Materials Instrumentation Facility (SMIF), a member of the North Carolina Research Triangle Nanotechnology Network (RTNN), which is supported by the National Science Foundation (Grant ECCS-1542015) as part of the National Nanotechnology Coordinated Infrastructure (NNCI). I.V.B. acknowledges CityU New Research Initiatives/Infrastructure Support from Central (APRC): 9610395, and the Hong Kong Research Grants Council (ECS) Projects: (ECS) 2301818, (GRF) 11303619.
\end{acknowledgments}

\nocite{*}

\bibliography{apssamp}% Produces the bibliography via BibTeX.
\bibliographystyle{apsrev4-2}

\section{Supplementary Information}

\renewcommand{\thepage}{S\arabic{page}}
\renewcommand{\thefigure}{S\arabic{figure}}
\renewcommand{\thesection}{S\arabic{section}}
\setcounter{figure}{0}
\setcounter{page}{1}
\setcounter{equation}{0}

\subsection{Fabrication}
Graphene and boron nitride flakes are separately exfoliated on a thermally oxidized silicon substrate. The graphene is then encapsulated between BN layers using a dry-transfer method, and deposited on a doped silicon substrate with a 280 nm thick oxide. The structure is then annealed in atmosphere at 500ºC for one hour. The device region is defined using electron beam lithography and is etched using a CHF$_3$ / O$_2$ reactive ion etching process. Superconducting electrodes consist of 70 nm thick molybdenum rhenium alloy which is sputtered directly after a reactive ion etch. The MoRe is sputtered at 70W in an argon atmosphere at a pressure of 3 mTorr.

\subsection{Simulation}

We consider a Josephson junction network with $N+1$ terminals. One of the contacts is grounded, so its  phase is assumed to be 0; there are therefore $N$ independent superconducting phases. This network contains $\binom{N}{2}$ Josephson junctions between each possible pair of terminals. We call $C_{j,k}$ and $G_{j,k}$ the capacitance and shunting conductance between terminals j and k, while $C_{j,j}$ and $G_{j,j}$ are by definition the capacitances and conductances to the ground terminal.

With these notations, the current flowing between terminals $j$ and $k$ due to one Josephson junction and its shunting capacitance and resistance is: 
\begin{equation}
    I_{jk}\sin(\phi_{j}-\phi_{k})+\frac{\hbar}{2e}G_{ij}(\dot\phi_{j}-\dot\phi_{k})+\frac{\hbar}{2e}C_{ij}(\ddot\phi_{j}-\ddot\phi_{k})
\end{equation}

Applying Kirchhoff's laws at each node results in N differential equations that can be written as:

\begin{multline*}
%\begin{aligned}
\frac{\hbar}{2e}\mathcal{C} \ddot\Phi +\frac{\hbar}{2e}\mathcal{G}\dot\Phi + f(\Phi)=G_{L}(V-\frac{\hbar}{2e}\dot\Phi) \\
C_{L}\dot V +G_{L} (V-\frac{\hbar}{2e}\dot\Phi)=I, \hspace{1mm} \\ \mbox{with}\hspace{1mm} \Phi=\begin{pmatrix} \phi_{1} \\ ... \\ \phi_{N} \end{pmatrix} \hspace{1mm} \mbox{,} \hspace{1mm} I=\begin{pmatrix} I_{1} \\ ... \\ I_{N} \end{pmatrix} \mbox{, and } V=\begin{pmatrix} V_{1} \\ ... \\ V_{N} \end{pmatrix}
%\end{aligned}
\end{multline*}

 $\mathcal{C}$ and $\mathcal{G}$ are NxN matrices whose coefficients depend on the capacitances and conductances respectively:
 
\begin{equation}
    \mathcal{C} = \begin{pmatrix}\sum_{k=1}^{N}C_{1,k} & ... & ... & ... & -C_{N,N} \\ ... & ... & ... & ... & ... \\  -C_{i,1} & ... & \sum_{k=1}^{N}C_{i,k} & ...&-C_{i,N} \\ ... & ... & ... & ... & ...\\ -C_{N,1} & ...&...&... & \sum_{k=1}^{N}C_{N,k} \end{pmatrix}
\end{equation}
\begin{equation}
    \mathcal{G} = \begin{pmatrix}\sum_{k=1}^{N}G_{1,k} & ... & ... & ... & -G_{N,N} \\ ... & ... & ... & ... & ... \\  -G_{i,1} & ... & \sum_{k=1}^{N}G_{i,k} & ...&-G_{i,N} \\ ... & ... & ... & ... & ...\\ -G_{N,1} & ...&...&... & \sum_{k=1}^{N}G_{N,k} \end{pmatrix}
\end{equation}

 Assuming a sinusoidal CPR for all junctions, $f(\Phi)$ is a N-row vector given by: 
 
\begin{equation}
    f(\Phi)= 
    \begin{pmatrix} 
    ... \\ I_{jj}\sin(\phi_{j})+\sum_{k\neq j}I_{jk}  \sin(\phi_{j}-\phi_{k}) \\
    ...
    \end{pmatrix}
\end{equation}

Here we labeled $I_{jk}$ the critical current of the Josephson junction between terminals j and k, while $I_{jj}$ refers to the Josephson junction from terminal j to ground.
This system of differential equations is equivalent to the following first order equations:

\begin{equation}
\begin{gathered}
%\begin{aligned}
\dot\Phi = \Psi \\
\dot \Psi = -\mathcal{C}^{-1}\mathcal{G} \Psi +G_{L}\mathcal{C}^{-1}(\frac{2e}{\hbar}V-\Psi)-\frac{2e}{\hbar}\mathcal{C}^{-1}f(\Phi) \\
C_{L}\dot V +G_{L} (V-\frac{\hbar}{2e}\dot\Phi)=I
\end{gathered}
\end{equation}

We solve this system of differential equations numerically using a fourth order Runge-Kutta routine.
\\
\\
\begin{table}
\begin{tabular}{|l|l|}
\hline
                                & Figure 4\\ \hline
$I_{c,LB}$ (nA)                 & 600      \\ \hline
$I_{c,RB}$ (nA)                  & 1100        \\ \hline
$I_{c,LR}$ (nA)                  & 500        \\ \hline
$R_{1}$ ($\Omega$)                  & 28.9         \\ \hline
$R_2$ ($\Omega$)                      & 13      \\ \hline
$R_3$ ($\Omega$)                      & 17.3      \\ \hline
$R_L$ ($\Omega$)                      & 50          \\ \hline
$C_{jj}$ (pF)                  & 1       \\ \hline
$C_L$ (pF)                       & 5   \\ \hline
$\omega$ (1/s)     & 3.267e10 \\ \hline
$I_{ac}$ ($\mu$A) & 17    \\ \hline
\end{tabular}
\caption{A list of parameters used in the main text.}
\end{table}

In the specific case of a three terminal Josephson junction, our above equations reduce to: 

\begin{equation}
%\begin{aligned}
\frac{\hbar}{2e}C \ddot\Phi +\frac{\hbar}{2e}G\dot\Phi + I_c(\Phi)=G_{l}(V-\frac{\hbar}{2e}\dot\Phi)
\end{equation}

\begin{equation}
C_{l}\dot V +G_{l} (V-\frac{\hbar}{2e}\dot\Phi)=I, \hspace{1mm}
\end{equation}

\begin{multline*}
\mbox{with}\hspace{1mm} \Phi=\begin{pmatrix} \phi_{L} \\ \phi_{R} \end{pmatrix} \hspace{1mm} \mbox{,} \hspace{1mm} I=\begin{pmatrix} I_{L} \\ I_{R} \end{pmatrix} \mbox{, and } V=\begin{pmatrix} V_{L} \\ V_{R} \end{pmatrix}
%\end{aligned}
\end{multline*}

Here, $\bf{C}$ and $\bf{G}$ are 2x2 matrices whose coefficients respectively depend on the three junctions capacitances, and the three conductances of the junctions' shunting resistors. $C_{l}$ and $G_{l}$ are the lead capacitance and conductance to the ground. For simplicity, the leads' parameters are taken to be the same for all contacts.  $I_c(\Phi)$ is a 2 row vector whose components are determined by the junctions' CPR and are functions of $\phi_{L}$, and $\phi_{R}$.

We use a fourth order Runge-Kutta method to solve $\Phi(t)$ and $V(t)$ over 100 microwave cycles. In order to determine the DC voltage at a given terminal, we time-average $V(t)$ over the last 90 microwave cycles. We take $I_c(\Phi)$ to be sinusoidal. Experimentally, the left contact is most strongly coupled to the drive and so we take the applied microwave power through the left and right leads to be different by 10 dBm. Additionally, an arbitrary phase offset is added between the two drives. The phase offset appears to be important in order to reproduce the observed fractional steps, which appear to be suppressed when the two AC currents are in phase. The full list of parameters used for our simulations and more detailed description of the program can be found in Table 1.

\subsection{Heating}

\begin{figure}[ht]
    \centering
    \includegraphics[width= \columnwidth]{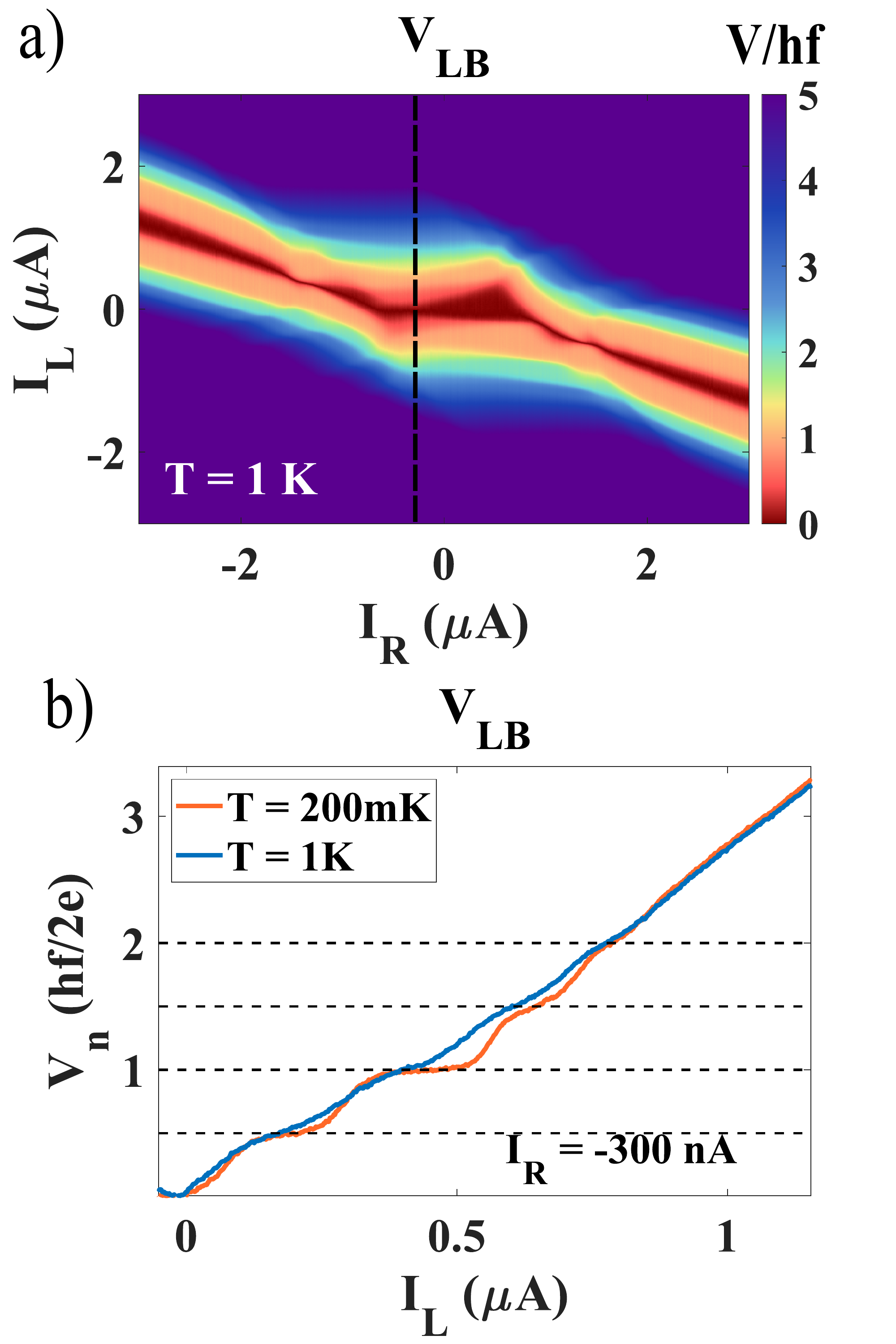}
 \caption { a) V$_{LB}$ bias-bias map measured using the same parameters as Figure 2a but at elevated temperature of 1 K. b) Comparison of the cut taken at $I_R$=-300 nA at 200 mK and 1 K. The degree of thermal suppression is similar for the 1/2 and 1 plateaus. Since higher harmonics of the CPR should be strongly suppressed by temperature, we conclude that the half-plateaus are not explained by the non-sinusoidal CPR.}
    \label{fig:Heating}
\end{figure} 

In some of the previous studies, fractional Shapiro steps were attributed to nonsinusoidality of the CPR. In that case, the fractional steps decayed with temperature faster than integer steps~\cite{InAs}. Indeed, the higher harmonics of the CPR should be strongly suppressed by temperature. In figure S1, we show that in our case, fractional steps demonstrate a similar degree of thermal suppression as the integer steps. This provides yet another strong piece of evidence that the fractional steps originate from the Josephson network.

\subsection{Hysteretic Regime}

The highly tunable nature of graphene Josephson junctions allows us to move between different dynamical regimes. Our devices have two primary knobs for varying dynamical regimes; critical current (I$_c$) and normal resistance (R$_N$). The critical current of the device dictates width of the Shapiro step in the Bessel function regime as: $I_c J_c (\frac{2eV_{AC}}{\hbar \omega})$. The critical current can be tuned, as it was in the main text, by varying the applied magnetic field. Meanwhile, the center of the Shapiro step in bias is dictated by $\hbar \omega / R_N$. The normal resistance can be tuned by applying a backgate voltage. 

At low I$_c$ and R$_N$, the Shapiro steps follow Bessel-like behavior and are well separated, allowing the full spectrum of steps to emerge. However, as either parameter is increased, the steps begin to overlap causing a competition between voltage values. When this happens, more dynamically stable steps dominate the weaker steps - washing them away. 

We demonstrate these effects by beginning with the power bias map produced in Figure 3b (Reproduced in Figure S2a). Here, a small magnetic field offset of .5 mT reduces the critical current allowing for the well separated Bessel functions to yield fractional steps. Next, we zero the applied magnetic field, while simultaneously decreasing backgate voltage. The net result fixes the critical current, while increasing the junction's normal resistance. The Shapiro steps deviate from Bessel behavior and only integer steps remain (Figure S2b). Consistent with the explanation in the main text, this demonstrates the integer steps are more dynamically stable than the fractional steps.

\begin{figure}[ht]
    \centering
    \includegraphics[width= \columnwidth]{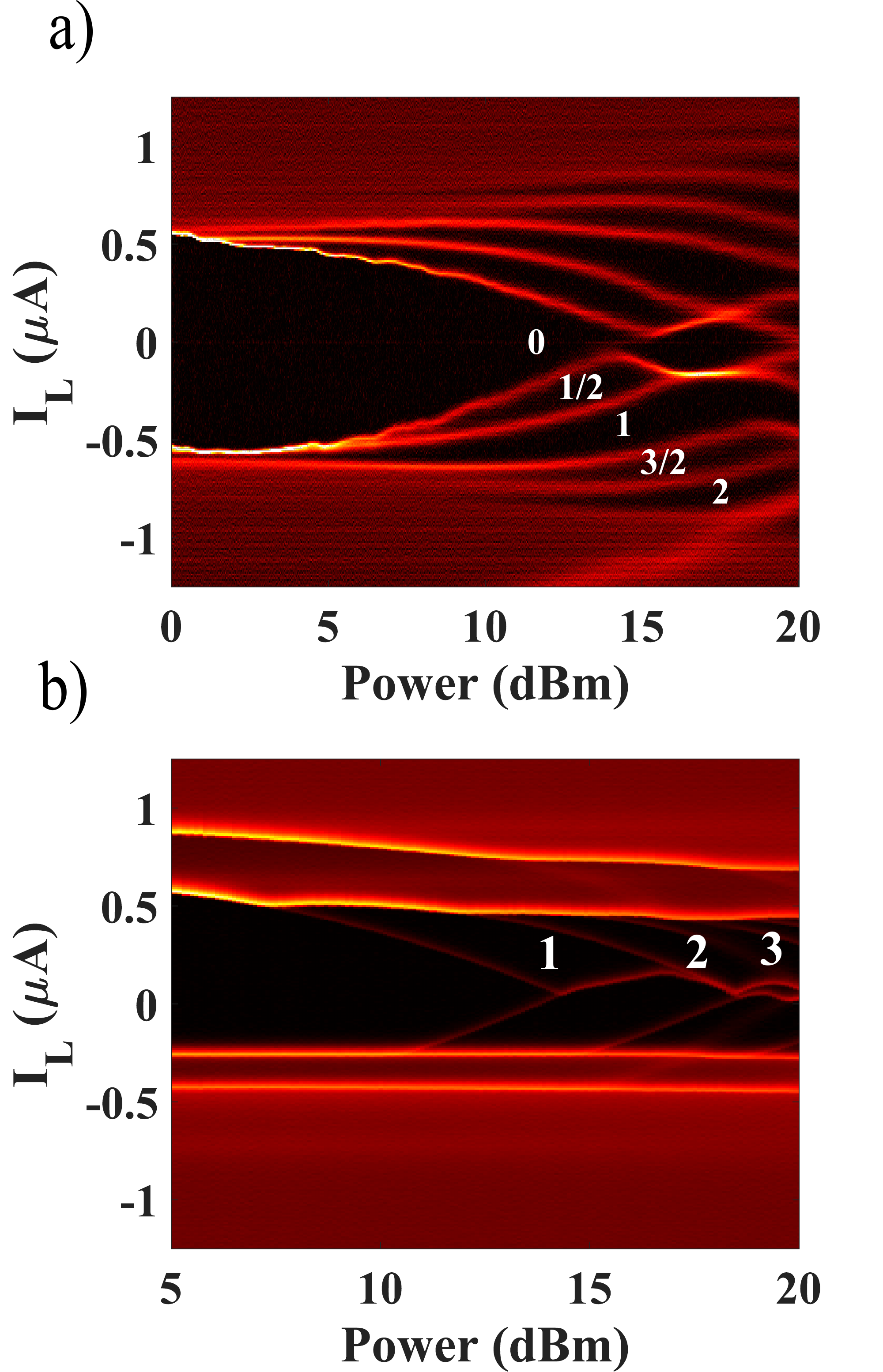}
 \caption {Comparison between different dynamical regimes. a) The Bessel regime, where an external magnetic field suppresses the critical current, while high backgate reduces R$_N$. In this case, the Shapiro steps are well separated allowing for the full spectrum of plateaus to emerge. b) The hysteretic regime, where a reduction in magnetic field and decrease in backgate cause the Shapiro steps to overlap. Here, the more dynamically stable integer steps overwhelm the fractional steps. This is consistent with the model presented in the main text.
    }
    \label{fig:Hysteretic}
\end{figure}

\subsection{Additional Frequencies}

To compliment the main text, we provide additional bias-bias voltage maps in a rotated geometry ({\bf L} is grounded, while current is applied from {\bf R, B}) at different frequencies. Fractional steps could be observed between any pair of junctions, depending of the relative coupling strength of the microwave excitation at a given frequency. Notably, at 4.73 GHz all three junctions display fractional steps within a given map. 

\begin{figure*}[ht]
    \centering
    \includegraphics[width= 2\columnwidth]{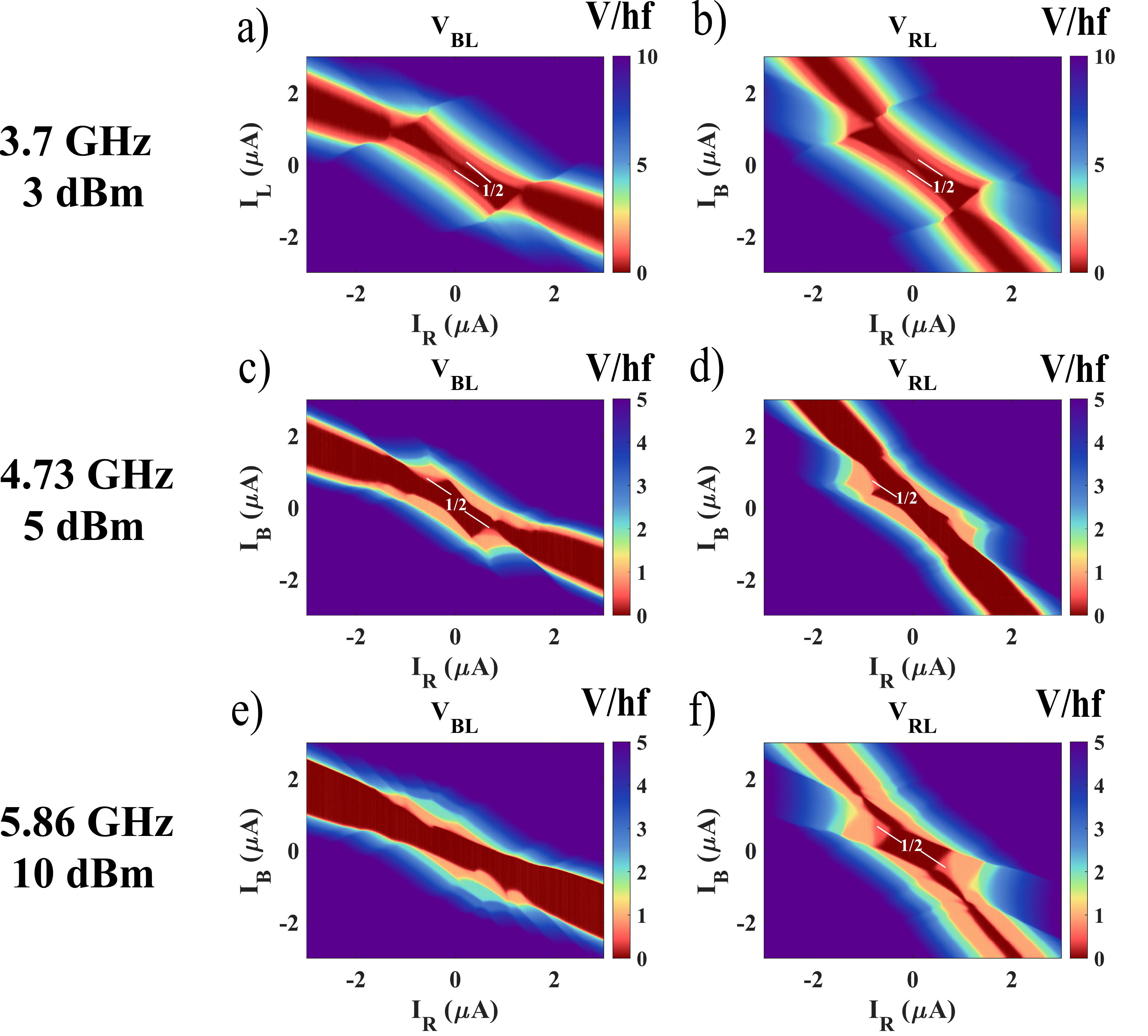}
 \caption {Bias-bias voltage maps spanning several frequencies. The biasing scheme is rotated as an additional check. Here, \textbf{L} is grounded, while current is applied from \textbf{R} and \textbf{B}. Fractional steps are observed in any pair of junctions. Notably, for 4.73 GHz (c,d), half steps are observed in all three junctions.
    }
    \label{fig:More}
\end{figure*}

\end{document}